\pdfoutput=1

\PassOptionsToPackage{table,xcdraw}{xcolor}
\documentclass[sigconf]{acmart}

\AtBeginDocument{%
  \providecommand\BibTeX{{%
    \normalfont B\kern-0.5em{\scshape i\kern-0.25em b}\kern-0.8em\TeX}}}

\setcopyright{acmcopyright}
\copyrightyear{2021}
\acmYear{2021}

\usepackage{booktabs}  
\usepackage{multirow}
\usepackage{tabularx}
\usepackage{paralist}
\usepackage{enumitem}

\acmSubmissionID{8212}

\begin{document}

\title{AniVis: Generating Animated Transitions Between Statistical Charts with a Tree-Based Model}


\author{Wenchao Li}
\email{wlibs@connect.ust.hk}
\affiliation{%
  \institution{The Hong Kong University of Science and Technology}
  \city{Hong Kong}
  \country{China}
}

\author{Yun Wang}
\affiliation{%
  \institution{Microsoft Research Asia}
  \city{Beijing}
  \country{China}}

\author{He Huang}
\affiliation{%
  \institution{Microsoft Research Asia}
  \city{Beijing}
  \country{China}}

\author{Weiwei Cui}
\affiliation{%
  \institution{Microsoft Research Asia}
  \city{Beijing}
  \country{China}}

\author{Haidong Zhang}
\affiliation{%
  \institution{Microsoft Research Asia}
  \city{Beijing}
  \country{China}}

\author{Huamin Qu}
\affiliation{%
  \institution{The Hong Kong University of Science and Technology}
  \city{Hong Kong}
  \country{China}}

\author{Dongmei Zhang}
\affiliation{%
  \institution{Microsoft Research Asia}
  \city{Beijing}
  \country{China}}

\renewcommand{\shortauthors}{Li et al.}

\begin{abstract}
Animated transitions help viewers understand changes between related visualizations. To clearly present the underlying relations between statistical charts, animation authors need to have a high level of expertise and a considerable amount of time to describe the relations with reasonable animation stages. 
We present AniVis, an automated approach for generating animated transitions to demonstrate the changes between two statistical charts. AniVis models each statistical chart on the basis of a tree-based structure. Given an input chart pair, the differences of data and visual properties of the chart pair are formalized as atomic transition units. Through this approach, the animated transition between two charts is expressed as a series of transition units. Then, we conduct a preliminary study to understand people's preferences for animation sequences. Based on the study, we propose a set of principles and a sequence composition algorithm to compose the transition units into a meaningful animation sequence. Finally, we synthesize these units together to deliver a smooth and intuitive animated transition between charts. 
To evaluate our workflow, we implement a web-based system and demonstrate its generated results to illustrate the usage of our approach. We perform a comparative study with the latest method to assess the animated transitions automatically generated by our system. We further collect feedback from experts to evaluate the usability of our method. 
\end{abstract}



\maketitle

\section{Introduction}
Many data visualizations are organized in sequences to tell data stories~\cite{hullman2013deeper, kim2017graphscape, segel2010narrative}. To completely perceive the implications of data, visualization viewers must examine not only individual charts but also the relationships between them~\cite{kim2017graphscape}. Animated transitions have been widely adopted between consecutive visualizations~\cite{heer2007animated, segel2010narrative, amini2016authoring}. Compared with static transitions, animated transitions are more effective in revealing data relationships, conveying semantic changes, supporting narratives, and preserving viewer engagement~\cite{chevalier2016animations, fisher2010animation, heer2007animated, amini2018hooked}. Users can thus easily understand data from multiple perspectives and follow narratives in a smooth and progressive way with less perception burden.

Nevertheless, without clear planning of transition stages, it still becomes difficult for the audience to follow the logic when complex changes happen in an animated transition~\cite{heer2007animated, fisher2010animation, bach2013graphdiaries, kim2019designing}. Consider an animated transition drill-down operation in a bar chart that includes the filtering of irrelevant items and visualizing an additional dimension. If all the changes are simultaneously animated with a fading effect, people might not be able to easily track and understand the shifting of the data information. In fact, the drill-down transition can be further broken down into transition stages of removing irrelevant bars, morphing the specified rectangular bar to a circular pie, and visualizing a new dimension on the pie, which provides a more effective data change presentation and helps the viewer better understand the animated transition between the source and the target charts. 

However, designing and authoring proper animated transitions to reveal the underlying data relations between data charts remains a challenging and time-consuming task. Animation authors must have a high level of expertise to identify the changes behind and a considerable amount of time to describe the data and the corresponding visual changes with reasonable animation stages. 
Moreover, they need to plan carefully and implement every animation steps, which requires abundant design experience and understanding in data visualization and a full mastery of professional design tools, such as Adobe After Effects and Cinema 4D or programming libraries, such as D3~\cite{bostock2011d3}. All the above obstacles make it hard for non-experts to generate the animated transitions for presenting data transformations in a short time. 

\begin{figure*}[t]
\centering
  \includegraphics[width=\linewidth]{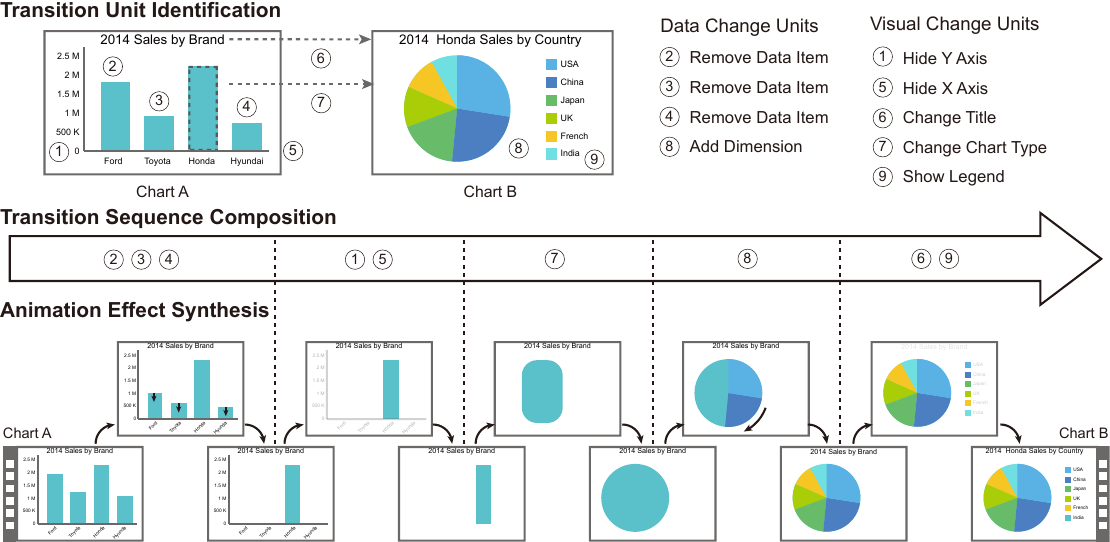}
  \caption{~\label{fig:pipeline}
  Overview of our method for the auto-generation of animated transitions: The pipeline consists of three major phases, namely, transition unit identification, transition sequence composition, and animation effect synthesis.}
\end{figure*}

Although prior research has either derived design guidelines and principles through user studies~\cite{tversky2002animation, dragicevic2011temporal, chevalier2014not, chalbi2019common} or designed and developed animation effects to support transitions between specific visualizations in an ad-hoc manner~\cite{heer2007animated, bach2013graphdiaries,  du2015trajectory, wang2017vector, kim2019designing}, there are few automated approaches for formulating the design process of animated transitions to convey data relations between general statistical charts. 
Recent work for specifying data graphics animations with high-level declarative languages~\cite{ge2020canis, kim2020gemini} provides a simpler grammar than D3~\cite{bostock2011d3} and gganimate~\cite{gganimate} for the user. However, they still require sufficient time and expertise to work with professional grammars. Kim and Heer~\cite{kim2020gemini} build a recommender for the transition steps by enumerating the transitions to reflect high-level changes of components (e.g., marks, axes, and legends). Then the recommender ranks the transitions by a heuristic edit operation cost functions tuned by a crowdsourced study, which may result in suboptimal single-stage animated transition designs that cannot clearly explain data relations between charts. 
The challenge of identifying and sequencing the detailed data changes while improving the understandability of animated transitions remains. 
Moreover, visual adjustments come with data changes (e.g., axis rescaling after the data filtering), prompting the inevitable question of how animated transitions for different changes can be systematically arranged. 

To reduce the efforts involved in identifying the underlying data relations and orchestrating both the data and the visual change process, we explore a new automated approach to systematically generate animated transitions between common statistical charts (i.e., bar charts, pie charts, line charts, and scatter plots). Developing such an approach is challenging because it needs to understand the data behind charts, recognize differences and connections of the graphical elements in charts, and identify various types of data and visual changes possibly involved in the transitions. To address the challenges, we first study the data and visual changes between any two data charts. We then model data as well as visual transitions between any two statistical charts as a sequence of transition units. 
Based on the transition model, we contribute AniVis, an automatic animated transition generation approach consisting of three modules, namely, transition unit identification, transition sequence composition, and animation effect synthesis (Fig.~\ref{fig:pipeline}). 

Given two statistical chart specifications as input, we first adopt a tree-based model to systematically represent the data of the charts, and then identify differences of both tree structures and visual properties between the two charts as transition units. For such adoption and identification, we differentiate the models of the charts and enumerate the visual edit operations. In addition, we conduct a preliminary study to understand people's preferences for data transition sequences. On the basis of the study results, we propose a set of principles and a transition sequence composition algorithm to arrange the transition units appropriately for producing a sequence of staged animations. Finally, we synthesize the animated transitions with a set of animation effects for the transition units to create the outcome animations. The generated animated transitions are designed to ease the tracking of visual changes and facilitate the understanding of data changes between charts for the audience. 

To understand how users interpret the animated transitions derived from our tree-based data model, we then perform a user study to compare with the transitions created by the latest method~\cite{kim2020gemini} and designers. In this study, we ask subjects to complete a ranking task on the basis of how easy the identification and understanding of the specific data changes in the transitions are. The ranking results, including the qualitative responses, indicate that our staged animated transitions that reveal the data change process at a more fine-grained level can improve the understandability of animated transitions between statistical charts. 
In addition, we validate the usability of our approach through semi-structured interviews with four experts. According to the feedback, the system offers a convenient way to explain the data changes between two visualizations with staged animations without writing code, and our approach can be used for the rapid drafting of user-desired animations.

\section{Related Work}
\subsection{Understanding Animation in Visualization}
Chevalier et al.~\cite{chevalier2016animations} summarize different roles of animation in a visual interface by extending the taxonomy proposed by Baecker and Small~\cite{baecker1990animation}. In particular, animations in visualization are used to reveal data relationships, support narratives, and keep the user engaged~\cite{chevalier2016animations,kale2018hypothetical,fisher2010animation}. Many studies have been conducted to understand the effectiveness of animation for data analysis and visualization. Although animation might be considered problematic in some tasks~\cite{tversky2002animation,  robertson2008effectiveness, fisher2010animation}, researchers have started to reveal the benefits of animation from a variety of perspectives. For example, Heer and Robertson~\cite{heer2007animated} show that animated transitions, when appropriately applied, can improve the graphical perception of statistical visualizations. Amini et al.~\cite{amini2018hooked} assess the effect of animation on viewer engagement of data videos and confirm that incorporation of animation can boost the understandability of data insights, and significantly improves viewer engagement. More recently, Ondov et al.~\cite{ondov2019face} have found that animated transitions perform better than static small-multiple arrangements in facilitating visual comparisons. These studies are orthogonal to our work as they attempt to understand and demonstrate the value of animation for visualizations, while our goal is to lower barriers for animation implementation. Specifically, we seek to automatically generate appropriate animated transitions between statistical visualizations. 

\subsection{Animated Transition Design}
Animated transition techniques have been employed in various data analysis and visualization tasks to preserve the context between data views. Specific animated transition designs have been proposed to reduce the cognitive burden for data exploration of complicated datasets (e.g., ScatterDice~\cite{elmqvist2008rolling} for multidimensional data, GraphDice~\cite{bezerianos2010graphdice} for multivariate social network data, and Matrix Cubes~\cite{bach2014visualizing} for dynamic networks), viewport switching~\cite{van2003smooth}, text transformation~\cite{chevalier2010using, dragicevic2011gliimpse}, storytelling with timelines~\cite{brehmer2016timelines}, and so on. A particular set of studies investigate animated transitions that can be utilized to depict various data analysis semantics related to statistical visualizations. Heer and Robertson~\cite{heer2007animated} present a taxonomy of transition types between statistical data graphics. They also provide design guidelines based on the Congruence and Apprehension principles of effective animation~\cite{tversky2002animation}. 
Visual sedimentation~\cite{huron2013visual} employs the metaphor of a physical sedimentation process to illustrate the aggregate operation that accumulates the values of incoming data streams. 
More recently, Kim et al.~\cite{kim2019designing} present and evaluate different staged animated transition designs to convey eight aggregate operations common to visual data analysis, instead of using linearly interpolated transitions. We extend the prior work along this direction by contributing a model to systematically represent transitions between common statistical charts. Based on the model, we build a general framework to automatically generate intuitive animations between charts. 



A single-step animation that moves all elements at the same time is usually insufficient to reveal complex changes during a visualization transition. Instead, a staged animation~\cite{heer2007animated}, which breaks down an animation into multiple steps, can help viewers easily follow visual transitions and better apprehend rich types of changes involved. Researchers have explored various aspects of staging design for animated transitions. For example, Chalbi et al.~\cite{chalbi2019common} compare different grouping strategies. Du et al.~\cite{du2015trajectory} propose a trajectory bundling approach for a group of adjacent objects that move in a similar direction. In addition to bundling the linear movement of points, Wang et al.~\cite{wang2017vector} propose a framework for creating animated transitions of points along nonlinear paths with collision avoidance. Dragicevic et al.~\cite{dragicevic2011temporal} perform an empirical study to evaluate the effectiveness of the animations with different paces. Many guidelines and insights derived from these studies inform the design of staged animations in our work. However, these techniques are isolated or optimized for one or more specific animation effects. Compared with them, we provide a systematic and integrated solution to automatically generate staged animations for conveying data changes between two statistical visualizations effectively.

\subsection{Chart Animation Authoring}
Recently, increasing efforts have been devoted to authoring chart animations~\cite{ge2020canis,kim2020gemini,thompson2021data,ge2021cast}. For example, Ge et al.~\cite{ge2020canis} introduce a high-level grammar, called Canis, which enables declarative specifications of data-driven chart animations. Their approach requires professional knowledge of the languages and detailed planning of the transitions between the start and end states. 
Most notably, the latest work by Kim et al.~\cite{kim2020gemini} takes the above declarative grammar idea further by introducing a recommendation system with change detection, enumeration, and ranking components. 
The key distinction between our approach and theirs is that our transition generation for all visual components is driven by a finer granularity of data changes automatically detected from the data transformations of the source and target charts. Our approach defines more specific data change units and systematically elaborates on how the data changes from one schema to the others step by step, which extends the data change detection module and transition recommendation in Gemini. Thus, independent transitions for more specific types of data changes (e.g., filtering, sorting, and merging) can be generated with our method, which allows the sequencing for different data change transitions and the grouping for the specific ones. 
Another difference lies in the transition unit: our framework considers the smooth switch between different chart types, whereas their transitions recommended by the system do not cover the low-level shape interpolation of mark elements (e.g., morphing between bar and pie). 

Instead of creating chart animations with declarative grammars that require people to write code, Ge et al.~\cite{ge2021cast} extend their previous work and present an interactive system to animate the visual elements within a single chart. Thompson et al.~\cite{thompson2021data} present another interactive system for authoring animated transitions between two static visualizations. 
Both of the systems ease the creation process without programming. However, their work does not target the automated recommendation of animation design for data changes. In contrast, AniVis constructs a tree-based data model that corresponds to the marks between two different charts, and can then be used to generate transition steps to reveal the data changes, which does not necessarily require specifying the keyframes.

\subsection{Sequencing in Narrative Visualization}
An effective sequence of visualizations is essential for conveying a data-driven narrative successfully. Hullman et al.~\cite{hullman2013deeper} conduct an initial study to understand the narrative sequencing of visualizations and proposed a graph-driven approach for automatically identifying effective sequences in a set of visualizations. Kim et al.~\cite{kim2017graphscape} further propose GraphScape, a directed graph model to support automated reasoning about visualization similarity and sequencing. While GraphScape determines the optimized sequence for a set of individual visualizations, our work arranges the orders of multiple animation steps to generate a staged animation for the transition from one statistical chart to another. Actually, GraphScape and our work are complementary. After GraphScape is used to automatically sequence a set of charts, our technology can be employed to automatically generate animated transitions for each adjacent chart pairs in the sequence. 

\subsection{Automatic Generation of Visualizations}
To reduce the efforts involved in creating visualizations, research on automated methods for visualization generation continues to expand for years. Various visualization recommendation systems, such as ShowMe~\cite{mackinlay2007show}, Voyager~\cite{wongsuphasawat2015voyager}, Voyager 2~\cite{wongsuphasawat2017voyager}, DeepEye~\cite{luo2018deepeye}, Draco~\cite{moritz2018formalizing}, and VizML~\cite{hu2019vizml}, use rule-based algorithms or machine learning models to recommend appropriate visual designs for a given data pattern. Recently, Text-to-Viz~\cite{cui2019text} automatically converts a natural language statement about proportion-related statistics to a set of infographics with different styles and settings. DataShot~\cite{wang2019datashot} creates fact sheets automatically from tabular data. All these works target the generation of static visualizations. By contrast, our work provides an automated pipeline for generating animated visualizations to facilitate the chart-to-chart transition. 

Commercial tools such as Flourish~\cite{flourish} enable users to use pre-defined templates to create data visualizations and animations easily. However, the data schema of the charts is fixed, and the expressiveness is limited by templates provided for specific charts in these tools. 
DataClips by Animi et al.~\cite{amini2016authoring} adopts a semi-automated approach to ease the creation of data videos with a built-in data clip library, which provides templates of a few types of animated visualizations commonly used in data videos. Most data clips only support transitions within one data view. By contrast, our work allows the auto-generation of animated transitions across visualizations with different data schema. Our system automatically identifies transition units for both data and visual properties between statistical charts and composes transition units as a sequence to generate staged animations.

\section{AniVis Overview}
We propose AniVis, a systematic approach for automatically generating animated transitions that communicate the underlying data relations as well as the visual changes between common statistical charts: bar charts, pie charts, line charts, and scatter plots. We consider them because they are the most frequently used chart types~\cite{harper2017converting}. 
To achieve this goal, we first derive a model to systematically represent transitions between common statistical charts. In specific, we model the transition between two statistical charts as a sequence of transition units. Serving as the building blocks of chart-to-chart transitions, transition units are defined to represent and describe various data changes and visual changes we want to convey in a chart-to-chart transition. 
With carefully designed transition units, we can build up animated transitions for various changes between charts. Multiple transition units can be further organized, ordered, and grouped to form an animated transition sequence with appropriate planning of time. 

Building on top of the transition model, AniVis seeks to provide an automated system that can generate animations between statistical charts. To guide the overall design of AniVis, we identify the design consideration as follows:

\begin{compactitem}
\item \textbf{Facilitating the understanding of data changes.} The system should create informative animations conveying data changes and data relationships between charts. 

\item \textbf{Generating staged animations.} A single-step animation is not suitable for clarifying complex changes during a visualization transition\cite{fisher2010animation}. The system should generate animations with multiple stages to gradually and clearly reveal various data change semantics between visualizations.

\item \textbf{Providing the flexibility to support transition configuration.} 
The approach should enable automatic generation with default animation designs. Meanwhile, it should also be flexible for users to configure animation timing and animation effects to provide alternative designs. 
\end{compactitem}

Based on our design considerations, we design AniVis as an auto-generation approach for systematically creating animated transitions to reveal the underlying data relations and the visual difference between two common statistical visualizations, which is generally a pipeline with three phases (Fig.~\ref{fig:pipeline}): 

\begin{compactitem}
\item \textbf{Transition Unit Identification.} The system first analyzes the specifications of the input chart pair and extracts data changes derived from the tree-based data model as well as visual changes between the two charts. All the changes are identified as corresponding data-related and visual-related transition units. 

\item \textbf{Transition Sequence Composition.} After identifying the transition units, the system further applies grouping and ordering algorithms to orchestrate these transition units into a sequence representing a chart transition with multiple stages.

\item \textbf{Animation Effect Synthesis.} Based on the composed transition sequence, the system finally outputs a staged animation by implementing each transition unit in the sequence with specific animation effects.
\end{compactitem}

\begin{figure}[tb]
\centering
  \includegraphics[width=0.97\columnwidth]{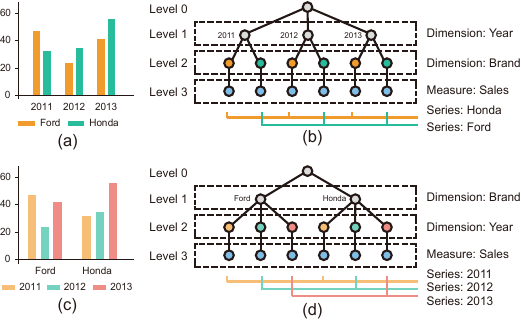}
  \caption{\label{fig:data_model}
  The tree-based data model of statistical charts: chart (a) and (c) are represented as tree (b) and (d), respectively. Note that the ordering of the dimensions in a tree depends on the data properties from the input specification.}
\end{figure}

\section{Transition Unit Identification with Tree Model}
\subsection{Tree-based Data Model Construction}
\label{section:tree_model_constuction}

In the context of data visualization for tabular data with the row and column structure, data columns are typically classified as dimensions and measures~\cite{mackinlay2007show}. Dimension columns are used to group or filter data records. The values in them are either categorical or temporal. Measure columns contain numerical values on which aggregations (e.g., \textit{sum} and \textit{average}) can be performed. Taking a car sales dataset as an example with four columns (i.e., ``Year'', ``Brand'', ``Country'', and ``Sales''), columns ``Year'', ``Brand'', and ``Country'' are dimensions, while column ``Sales'' is a measure. 

A statistical chart is a structured visual representation of measure values on a specific data subspace or multiple data subspaces, which can be achieved by slicing-and-dicing data records on dimension(s) and calculating measure(s) for each divided subspace against tabular dataset. For example, to generate a bar chart of ``sales by brand'', we need to first group the data records by the dimension ``Brand'', and then sum the values in measure ``Sales'' for each brand.

Considering sequential data preparation operations required for statistical charts, we build a tree model to represent data behind each chart (Fig.~\ref{fig:data_model}).
Intuitively, the data of a chart is a tree where the root node represents the whole tabular data. Each intermediate level of the tree corresponds to a dimension, where a node is a specific value of this dimension. The leaf nodes, corresponding to measure values, represent the data points that are visually plotted on the chart.
Leaf nodes are grouped into series based on the values of the lowest dimension.
Note that the tree-based model is independent of chart type, and our tree model is designed to represent statistical charts with aggregated values. In the special case of supporting the non-aggregated data points in scatter plots, we add an extra level to the leaves of the tree to represent the raw values. 

Given a chart, we can derive its tree model based on the visual encoding specification of the chart. For instance, Fig.~\ref{fig:data_model}(a) is a clustered bar chart with dimension ``Year'' mapped to X axis and ``Brand'' mapped to legend. To construct its tree model, we first add the dimension values of X axis to the tree, then expand the tree to add the dimension values on legend, and finally add the measure nodes (``Sales"). 

The order of the levels in the tree model reflects the grouping sequence of data points plotted on a chart. For two charts with a same set of data points, different orders of internal levels may result in different charts. For example, Fig.~\ref{fig:data_model}(a) and Fig.~\ref{fig:data_model}(c) are both clustered bar charts with same data points, but their grouping structures are different. In Fig.~\ref{fig:data_model}(a), the data points are first grouped by ``Year'' on X axis and are then grouped by ``Brand'' for each value of ``Year'', while in Fig.~\ref{fig:data_model}(b), the data points are grouped in a reverse order of ``Year'' and ``Brand''. Accordingly, the dimensions are placed on different levels of corresponding tree models (Fig.~\ref{fig:data_model}(b) and (d)).

\begin{figure*}[t]
\centering
  \includegraphics[width=0.97\linewidth]{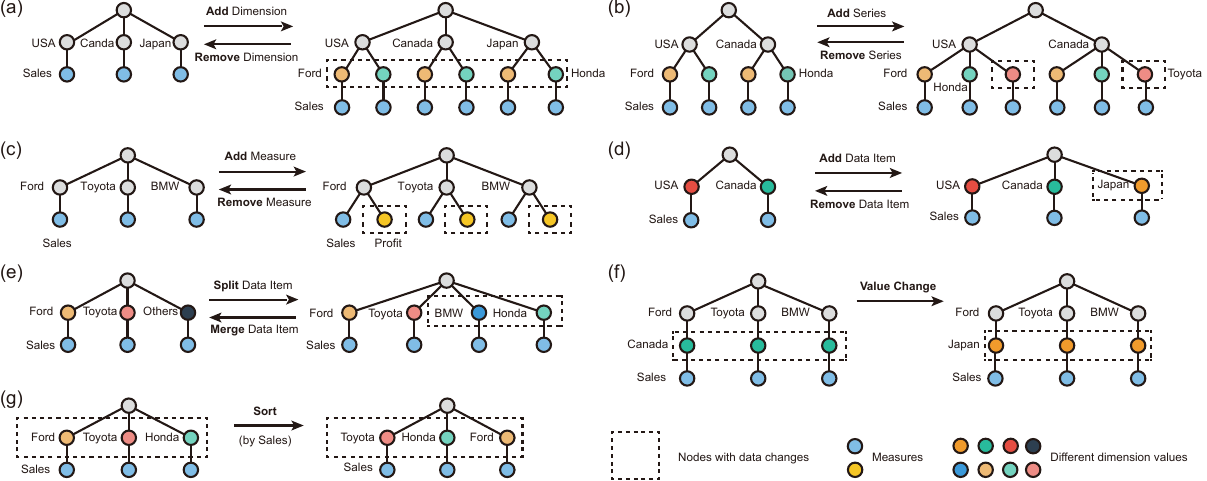}
  \caption{~\label{fig:data_change}
  Data change units that are identified from two statistical charts on the basis of the tree model: (a) Add/Remove Dimension, (b) Add/Remove Series, (c) Add/Remove Measure, (d) Add/Remove Data Item, (e) Split/Merge Data Item, (f) Value Change, and (g) Sort.}
\end{figure*}

\subsection{Transition Unit Identification}
We first analyze the data and visual changes between two statistical charts and model them as transition units. There are two types of transition units, namely data change units and visual change units, which correspond to data changes and visual changes during the transition, respectively. 

\subsubsection{Data Change Units}
We model chart data as a tree organized with four types of elements: dimensions, measures, series, and data items. Inspired by the previous taxonomies of visualization transitions~\cite{heer2007animated} and edit operations between visualizations~\cite{hullman2013deeper, kim2017graphscape}, we define a set of data change units to present data changes between two charts that corresponds to different tree edit operations on the aforementioned elements. Specifically, the data change units have 12 types (Fig.~\ref{fig:data_change}): \textit{Add Dimension}, \textit{Remove Dimension}, \textit{Add Measure}, \textit{Remove Measure}, \textit{Add Series}, \textit{Remove Series}, \textit{Add Data Item}, \textit{Remove Data Item}, \textit{Split Data Item}, \textit{Merge Data Item}, \textit{Sort}, and \textit{Value Change}. 
Each data change unit corresponds to a certain tree edit operation on the tree model of a statistical chart. For example, an \textit{Add Dimension} unit adds a new level to the tree hierarchy; a \textit{Remove Series} unit deletes a branch from the leaf level.

\begin{figure}[tb]
\centering
  \includegraphics[width=0.97\columnwidth]{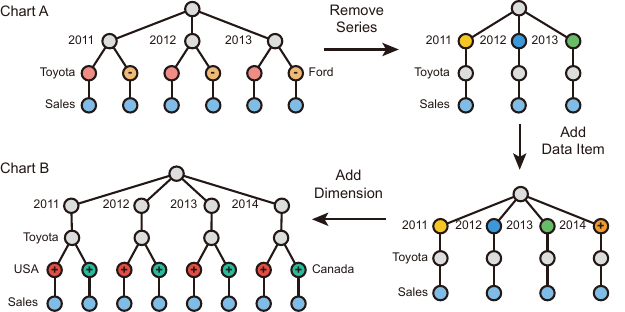}
  \caption{A transition between Chart A and Chart B with a sequence of data change units: ``-'' denotes the nodes to be removed, and ``+'' denotes the nodes to be added.}~\label{fig:sequence}
\end{figure}

On the basis of the tree model, we identify the data change units between two charts (e.g., Chart A and Chart B) by comparing the corresponding trees (e.g., \textit{Tree A} for Chart A and \textit{Tree B} for Chart B) to get a sequence of data change units (tree edit operations) that can transfer \textit{Tree A} to \textit{Tree B}, as illustrated in Fig.~\ref{fig:sequence}. To achieve this, we first locate the aligned level between \textit{Tree A} and \textit{Tree B}. The aligned level \textit{N} is determined if the corresponding dimensions of \textit{Tree A} and \textit{Tree B} are the same for each level from 0 to \textit{N}, but different for level \textit{$N + 1$}. Then, we conduct edit operations based on the aligned level to transfer \textit{Tree A} to \textit{Tree B}: 

\begin{compactenum}
\item \textbf{Node Removal.} If there are any data items or series at the aligned level of \textit{Tree A} but not in \textit{Tree B}, we remove them from \textit{Tree A};
\item \textbf{Subtree Removal.} If there are levels beneath the aligned level in \textit{Tree A}, we remove corresponding dimensions from \textit{Tree A} from the bottom up as the nodes at such levels are not in \textit{Tree B}; If the parent node and all its leaf nodes are to be removed, we remove the subtree directly to reduce redundant edits;
\item \textbf{Subtree Addition.} If there are levels beneath the aligned level in \textit{Tree B}, we will add corresponding dimensions to \textit{Tree A} from the top down.
\item \textbf{Node Addition.} If there are any data items or series at the aligned level of \textit{Tree B} but not in \textit{Tree A}, we add them to \textit{Tree A}. 
\end{compactenum}

With a sequence of tree edit operations, \textit{Tree A} will be transferred to \textit{Tree B}. We can then identify a sequence of corresponding data change units between two charts. If the generated tree models of the two charts are the same, the transition unit identification module will further compares the visual properties in their specifications (e.g., \textit{chartType}, \textit{isShowLegend}, and \textit{title}).

\subsubsection{Visual Change Units}
Visual aspects, such as chart type, axis, legend, and chart title, may vary between two statistical charts. We define a set of transition units corresponding to the transition operations related to visual changes. 
There are two types of visual change units, visual-related transition units and data-dependent transition units.

Visual-related transition units can be easily detected by comparison between chart specifications of the two charts, including
\textit{Show X Axis}, \textit{Hide X Axis}, \textit{Show Y Axis}, \textit{Hide Y Axis}, \textit{Show Legend}, \textit{Hide Legend}, \textit{Change Chart Type}, and \textit{Change Title}.
For example, a \textit{Change Chart Type} unit is identified for a transition between a pie chart and a line chart. A \textit{Hide Y Axis} is identified for a transition from a column chart with Y Axis to another column chart without Y Axis. A \textit{Change Title} is identified for a transition from two charts with different title descriptions. 

Data-dependent visual units are the visual change units generated from the data change units, including \textit{Rescale X Axis} depending on \textit{Add/Remove Series}, \textit{Add/Remove Data Item}, \textit{Merge/Split Data Item}, \textit{Add/Remove Dimension}, \textit{Rescale Y Axis} depending on \textit{Value Change}, \textit{Add/Remove Measure}, and \textit{Update Legend} depending on \textit{Add/Remove Dimension}, \textit{Add/Remove Series}, \textit{Add/Remove Measure}. These visual components change along with data operations.

\section{Transition Sequence Composition}
\label{sec:transition_sequence_composition}
To compose meaningful and effective animated transitions, we need to further arrange the transition units along the time axis. We consider the principles of staging \cite{heer2007animated} and concise \cite{fisher2010animation}, and compose the animated transitions in a three-step manner: ordering, grouping, and duration. 

\subsection {Formative Study}


To determine proper sequencing in our tree-based data model, we conducted an online experiment to understand how people choose among multiple possible transition sequences. We recruited participants from Prolific, a crowdsourcing platform, to select their preferences for the sequences of different data change units. The participants' task is to choose the appropriate transition process from the candidates for changing a source chart to a target one. 

\begin{figure}[tb]
\centering
  \includegraphics[width=0.97\columnwidth]{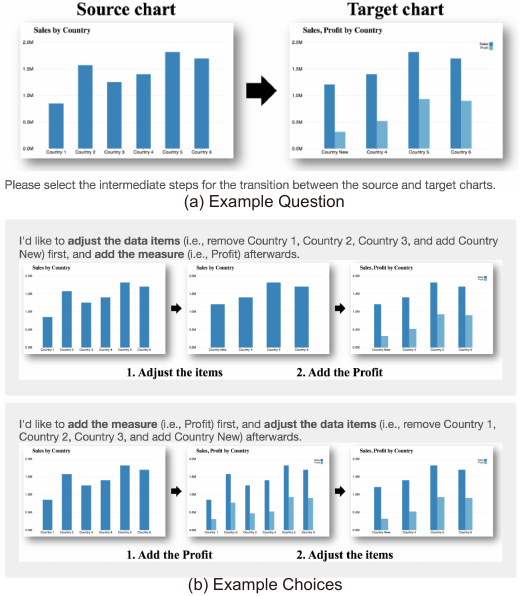}
  \caption{Example of the question and the corresponding choices in an experiment page: The question provides the source and target charts of an animated transition, including the changes involved. The choices are snapshots that represent different sequences of transition units.}~\label{fig:sequence_question}
\end{figure}

\subsubsection{Experiment Procedure}
Before starting the experiment, the participants were requested to read a tutorial introduction about the experiment settings and the possible transitions for data changes between two statistical charts. To help the participants recognize the difference between transition sequence, we provided a video to demonstrate two different transition sequence examples of the same chart pair. 
The participants were also suggested to consider the conflicts between operations and the number of edit elements for different sequences. 
Once the participants finished the tutorial, they were directed to the main experimental session and started to select the animated transitions that explains the data changes between two statistical charts clearly and logically.

We generated 12 chart pairs that varied the types of data visualization (vertical bar chart, horizontal bar chart, or line chart) with a synthesized car sales dataset. Each chart pair was constructed to test specific transition orders in a certain condition. The transition of chart pair covers the data change units of \textit{Add/Remove Data Item}, \textit{Add/Remove Dimension}, \textit{Add/Remove Series}, and \textit{Add/Remove Measure}. We omitted the data change units of \textit{Merge/Split Data Item}, \textit{Value Change}, and \textit{Sort} because their data operations on the tree model can essentially be replaced by the \textit{Add/Remove Data Item} unit. 
The chart pairs presented to the participants were used to assess the priority order of the data change units taking place in different levels of the tree model. 
Because the dimension, measure, and series cannot be simultaneously visualized in a common statistical chart, we selected the following combinations of (\textit{Add Data Item}, \textit{Remove Data Item}), (\textit{Adjust Data Item}, \textit{Add/Remove Dimension}), (\textit{Adjust Data Item}, \textit{Add/Remove Measure}), and (\textit{Adjust Data Item}, \textit{Add/Remove Series}) to investigate the preferred data change unit sequences take place within a single level of the tree model and those involve two levels. 
Note that we used the \textit{Adjust Data Item} to represent the combination of \textit{Add Data Item} and \textit{Remove Data Item} to let participants concentrate on the key difference between different types of data change units. 

As an example in Fig.~\ref{fig:sequence_question}, the participants picked a sequence from two candidates at a time. They were required to choose the order of the data operations (i.e., adding and removing the data items, and adding the profit measure) in a transition to determine the priority of the two types of data change units (i.e., \textit{Adjust Data Item} and \textit{Add/Remove Measure}) in the transition sequence. The participants were also asked to explain their choices. The occurrence of the chart pair and the order of the choices were randomized. All stimuli are available in the supplementary  material. 
Overall, 48 participants (12 females and 36 males; aged 18 - 43 [median = 23.7, SD = 5.56]) successfully completed the study and spent 24.3 minutes on average for the entire session. They were compensated at an average rate of 6.33 GBP/hour. 

\subsubsection{Results}
\begin{table}[tb]
\renewcommand{\arraystretch}{1}
\caption{Preferred order frequencies for different combinations of data change units from the participants.}
\label{table:sequence_result}
\centering
\begin{tabularx}{0.98\columnwidth}{p{0.2\columnwidth}<{\centering}p{0.33\columnwidth}p{0.33\columnwidth}}
\toprule
\textbf{Change} & \multicolumn{2}{c}{\textbf{Sequence}} \\ \hline
\textbf{\begin{tabular}[c]{@{}c@{}}Within\\ Dimension\end{tabular}}  & \begin{tabular}[c]{@{}l@{}}Add Data Item $\rightarrow$\\ Remove Data Item\\ \textbf{73/288 (25.3\%)}\end{tabular}                            & \begin{tabular}[c]{@{}l@{}}Remove Data Item $\rightarrow$\\ Add Data Item\\ \textbf{215/288 (74.7\%)}\end{tabular}      \\ \hline
\textbf{\begin{tabular}[c]{@{}c@{}}Dimension\\ Added\end{tabular}}   & \begin{tabular}[c]{@{}l@{}}Adjust Data Item $\rightarrow$\\ Add Dimension\\ \textbf{33/48 (68.8\%)}\end{tabular} & \begin{tabular}[c]{@{}l@{}}Add Dimension $\rightarrow$\\ Adjust Data Item\\ \textbf{15/48 (31.2\%)}\end{tabular}                            \\ \hline
\textbf{\begin{tabular}[c]{@{}c@{}}Dimension\\ Removed\end{tabular}} & \begin{tabular}[c]{@{}l@{}}Adjust Data Item $\rightarrow$\\ Remove Dimension\\ \textbf{9/48 (18.8\%)}\end{tabular}                       & \begin{tabular}[c]{@{}l@{}}Remove Dimension $\rightarrow$\\ Adjust Data Item\\ \textbf{39/48 (81.3\%)}\end{tabular} \\ \hline
\textbf{\begin{tabular}[c]{@{}c@{}}Measure\\ Added\end{tabular}}     & \begin{tabular}[c]{@{}l@{}}Adjust Data Item $\rightarrow$\\ Add Measure\\ \textbf{32/48 (66.7\%)}\end{tabular}   & \begin{tabular}[c]{@{}l@{}}Add Measure $\rightarrow$\\ Adjust Data Item\\ \textbf{16/48 (33.3\%)}\end{tabular}                              \\ \hline
\textbf{\begin{tabular}[c]{@{}c@{}}Measure\\ Removed\end{tabular}}   & \begin{tabular}[c]{@{}l@{}}Adjust Data Item $\rightarrow$\\ Remove Measure\\ \textbf{21/48 (43.8\%)}\end{tabular}                        & \begin{tabular}[c]{@{}l@{}}Remove Measure $\rightarrow$\\ Adjust Data Item\\ \textbf{27/48 (56.3\%)}\end{tabular}   \\ \hline
\textbf{\begin{tabular}[c]{@{}c@{}}Series\\ Added\end{tabular}}      & \begin{tabular}[c]{@{}l@{}}Adjust Data Item $\rightarrow$\\ Add Series\\ \textbf{36/48 (75.0\%)}\end{tabular}    & \begin{tabular}[c]{@{}l@{}}Add Series\\ Adjust Data Item\\ \textbf{12/48 (25.0\%)}\end{tabular}                                \\ \hline
\textbf{\begin{tabular}[c]{@{}c@{}}Series\\ Removed\end{tabular}}    & \begin{tabular}[c]{@{}l@{}}Adjust Data Item $\rightarrow$\\ Remove Series\\  \textbf{16/48 (33.3\%)}\end{tabular}                        & \begin{tabular}[c]{@{}l@{}}Remove Series $\rightarrow$ \\ Adjust Data Item\\ \textbf{32/48 (66.7\%)}\end{tabular}   \\ 
\bottomrule
\end{tabularx}
\end{table}

We collected a total of 576 choices for data change unit sequences (12 per participant). Table.~\ref{table:sequence_result} shows the numbers and the frequencies of the participants' preferred orders from animated transition creation perspectives. 

A reasonable expectation is that participants would prefer orders that minimize the number of edited visual elements across each chart pair. We designed trials (with different data abstraction levels and data visualization types) to test participants' preferences regarding adding versus removing data items within a dimension level. We found a clear majority direction for removing (74.7\%) before adding items (25.3\%). This is because removing before adding can reduce the overlapping of the visual elements to be adjusted. 

For the general-to-specific transitions (e.g., a dimension added, a measure added, or a series added), 70.1\% placed the data change units on the general data level first versus 29.9\% who increased the level of data summarization before adjusting the data items. The results support the strategy of starting the adjustment on a higher level of the tree-based data model and then splitting its nodes when we need to add an additional level to the tree. 

For the specific-to-general transitions (e.g., a dimension removed, a measure removed, or a series removed), 68.1\% adjusted the lower level of the tree model first versus 31.9\% who modified the items on a higher one. Still, the results largely agree with the expected tactic that minimizes the number of visual elements to be edited in a transition and support the default order of adjustment on a lower level of the tree before roll-up when the level removal occurs.

\subsection{Ordering}
On the basis of the results from the formative study, the tree model outputs an ordered sequence of data change units in the phase of transition unit identification. The phase also identifies a set of visual change units in correspondence to the data change units. Naturally, the data change units are followed closely by the corresponding data-dependent units to avoid the scale domain overflow of the visual marks' data~\cite{heer2007animated}. Thus, both the data-dependent units and the data change units ensure a meaningful sequence of transition unit groups~\cite{fisher2010animation}. 

To further compose the animation sequence, we must carefully insert the isolated visual change units into the sequence of data change units and output a complete transition. 
From the transition unit identification stage, the transition units can be divided into two parts, removing and adding graphical components. To minimize occlusion in the animation~\cite{heer2007animated}, we insert \textit{Change Chart Type} between these two stages. If no \textit{Change Chart Type} is identified, then we skip this animation.
To prepare chart type change, we should align the axes to prepare for the graphical components of the second chart to ensure the compatibility~\cite{fisher2010animation}. Therefore, we hide axes before \textit{Change Chart Type} and show axes after that.
Finally, we put the rest of the visual change units, including \textit{Show/Hide Legend} and \textit{Change Title}, at the end of the transition sequence. 

\begin{figure}[tb]
\centering
  \includegraphics[width=0.97\columnwidth]{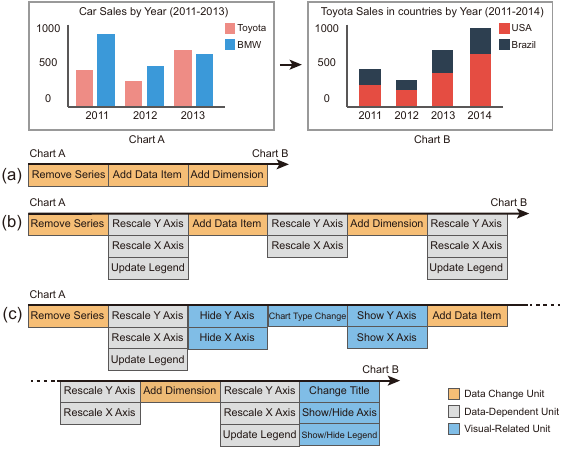}
  \caption{Composition of animated transition: (a) sequence of data change transition units; (b) add data-dependent units to the sequence; (c) insert visual-related units to the transition sequence.}~\label{fig:composition}
\end{figure}

Fig.~\ref{fig:composition} shows the generation of animated transition sequence from Chart A to Chart B. (a) From the transition identification phase, we have the sequence of data change units: \textit{Remove Series}, \textit{Add Data Item}, \textit{Add Dimension}. (b) We identify the data-dependent units related to the data changes. In this example, \textit{Rescale X Axis} for \textit{Remove Series}, \textit{Add Data Item}, and \textit{Add Dimension} is used to adjust the plot area. \textit{Update Legend} is inserted along with data changes. \textit{Rescale Y Axis} is not used because the y-axis is the same. (c) To complete the transition sequence, we insert the remaining visual-related units detected from the transition unit identification stage to the designated points of the sequence.

\subsection{Grouping}
To enhance the perception of the generated transition, we optimized the composition process by a grouping strategy based on the Gestalt Law of Common Fate~\cite{chalbi2019common, koffka1922perception}, where visual elements that move at the same pace tend to be recognized as a unified group. 
The well-known animation design principle facilitates the understanding that the visual elements are performing the same operation simultaneously and helps us structurally organize the data changes in the transition. By default, we group all the transition units of the same type together. For example, we merge all the data item removal transitions into one when irrelevant items need to be cleaned. Subsequently, the attention is directly guided, and anticipation is built for the remaining ones. 

In a staged animation, avoiding over-complex transition with too many stages~\cite{heer2007animated, kim2019designing} is also preferable. We further combine the minor transition units (i.e., legend and title-related units) into a single one to reduce the total number of stages, leading to a less complicated final transition. For example, we grouped the transitions units for updating the title and legend together as the final stage of the transition in Fig.~\ref{fig:pipeline}. The grouping strategy in the composition process shortens the animated transition to retain the engagement~\cite{heer2007animated} and helps viewers maintain focus on the visual marks, which are essential for representing the data changes of the animated transitions.

\subsection{Duration}
We also consider the temporal aspect of the final animated transition~\cite{thompson2020understanding}.
The duration has a significant influence on explaining the data changes. In AniVis, there are two alternative strategies for controlling the duration of the generated transitions as well as the transition units: (1) Animation-based duration and (2) fixed time duration. 

\textbf{Animation-based duration.}
A transition unit may be broken down into several animation effects. For example, the transition unit of \textit{Change Chart Type} can be divided into two effects of \textit{morphing} and \textit{move} graphical components, while the transition unit of \textit{Sort} contains only one animation effect. Inspired by the work of van Wijk and Nuij~\cite{van2003smooth} that matches the duration of a transition to the covered distance, we set the duration for each transition unit by depending on the ``animation steps'' in it. 
An animation step may be the appearance, exit, or update of a visual elements. As a result, the duration of each transition unit is the sum of the durations for corresponding animation effects. 

\textbf{Fixed-time duration.}
An alternative way to allocate the time duration is to set an overall duration for the final animated transition. The duration for transition units with different animation effects is equally distributed. For instance, we set a fixed duration (e.g., 2000 ms) for a whole animated transition between two input charts. 

In our default setting, we adopt the first strategy. We set 500 ms for the duration of each animation step (i.e., animation effect) in a transition unit. Each transition unit consists of 1 to n animation steps, resulting in a duration of $500 \times n$ ms. 
In addition, setting an appropriate standing time contributes to highlighting the key information, avoiding an overwhelming experience~\cite{kim2019designing}. We empirically set 1000 ms as the standing time before the data-related units, 500 ms before the mark change unit, and 0 ms for the animation steps within a transition unit.

\section{Animation Effect Synthesis}
\label{sec:animation_effect_synthesis}
Once we compose a sequence of transition units from the transition sequence composition phase, AniVis further synthesizes the final animated transition with specific visual effects for each type of transition units. The design of animation effects considers not only the meanings each transition unit conveys, but also the properties of different chart types. 

To choose and implement the default animation effect designs, we examine a collection of over 50 data videos provided by the previous work~\cite{amini2015understanding, amini2016authoring} as well as commonly used tools like Microsoft PowerPoint to obtain and analyze typical animation effects. 
Then, we follow the design considerations and principles for animation~\cite{heer2007animated,fisher2010animation}, such as using \textit{simple transitions} and \textit{meaningful motions}, and exemplify the transition units with commonly used visual effects, attempting to build a perception of the data changes in the final animated transition. We design the animations for the data change units and visual change units from the following three aspects: 

\begin{compactitem}
 \item \textbf{Entrance.} The entrance animations involves data change transitions such as \textit{Add Series}, \textit{Add Data Item}, \textit{Add Measure} and visual-related transitions such as \textit{Add X Axis}, \textit{Add Y Axis}. For visualization charts, it shows the appearance of graphical units on the chart areas. For example, a new bar for bar chart, or a series of points appear in scatter plots. For entrance, we can take animation effects such as \textit{wipe}, \textit{fly in}, or \textit{fade in}. 

 \item \textbf{Exit.} The exit animations mean graphical units disappear from the charts. They involve data change units such as \textit{Remove Series}, \textit{Remove Data Item}, and \textit{Remove Measure}. For exit, we can take animation effects such as \textit{shrink}, \textit{fly out}, or \textit{fade out}. 
 \item \textbf{Updates.} The updated animations show the changes of the graphical units, such as \textit{color change}, \textit{morph}, and \textit{move}. They are usually used for data change units such as \textit{Add/Remove Dimension}, \textit{Merge/Split Data Item}, \textit{Value Change}, and \textit{Sort}, and visual-related transition units such as \textit{Change Chart Type}.

\end{compactitem}

Since numerous animation designs can be adapted to show the semantic meaning for each transition unit, as a first step, we choose to implement one default effect for each transition unit for different chart types to enable the automation of animated transition generations. As AniVis is a general method of creating animated transitions, we do not include the animation effects for \textit{Emphasis}. More animation effects can be adopted in the future to extend the current animation effect library to support more customized animated transitions and improve the variability of transitions generated by AniVis.

For \textbf{data change transition units} in different chart types, we support 12 data change transitions for four types of charts (bar charts, line charts, pie charts, scatter plots). 
The animation effects for transition units of \textit{Add Series}, \textit{Add Data Item}, and \textit{Add Measure} usually involve \textit{entrance} effects. For example, we may grow the newly added bars to show new items, series, or measures in a bar chart. Correspondingly, the animation effects for transition units of \textit{Remove Series}, \textit{Remove Data Item}, and \textit{Remove Measure} usually involve \textit{exit} effects. For example, we may fade out nodes and the connecting lines to show removed items in a line chart. Transition units of \textit{Add/Remove Dimension}, \textit{Merge/Split Data Item} usually map to the \textit{update} effects. More specifically, for the chart types we support, add and remove a dimension are shown with the change of colors, while the merge and split of data items are represented with animations of \textit{move} and \textit{merge} or \textit{split}. In addition, \textit{Value Change} is updated according to the mark type, by changing the size, position, or length, while \textit{Sort} is a special type of animation, which only appears in the bar and pie charts and shown with position adjustments. 

For \textbf{visual change transition units}, we support showing and hiding legends and titles with a \textit{fade in} and \textit{fade out} effect. To support chart type changes, we implemented a total of $P(5,2)=20$ different animations covering the transformation of different chart types when the underlying data is aligned. More specifically, statistical charts supported in our system are encoded with mark types of area (vertical/horizontal bar chart, pie chart), length, and position (line chart, scatter plot). When mark change happens within the same mark types, we simply take \textit{morphing} and \textit{move} effects. When the mark changes between different mark types, we take more steps. We gradually morph the marks in a point-line-area order or vice versa. For example, when we change from a bar chart to a scatter plot, we gradually shrink the width of the bars, and then shrink the length into points, encoding the values with positions. After that, we add the axis and move the points to the correct positions (corresponding to \textit{Add Measure} as a data-related transition unit). The only exception is the animation for \textit{mark change} between a line chart and scatter plot (with point as mark) and pie chart. Following our principle, we need to extend points to lines, expand the line into rectangles, and morph rectangles into circles. This design can effectively convey the meaning of operations. However, we take a short cut to directly morph points into circles to avoid a too long and tedious animation process. 


\section{System Interface}
\begin{figure}[t]
\centering
  \includegraphics[width=0.97\columnwidth]{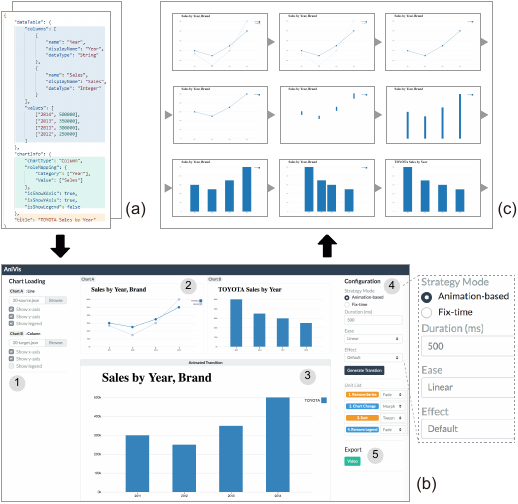}
  \caption{\label{fig:poc}
  The system workflow: The user loads the input chart pair via JSON files (a). The user can modify chart settings in the left panel (b1), and charts are rendered in the preview panel (b2). An animated transition between the two charts is automatically generated by AniVis and is displayed in the animation canvas (b3). The user can further refine the animation via the transition configuration panel (b4). Finally, by clicking the ``Export'' button (b5), users can save the animation results as an MP4 or GIF file (c).}
\end{figure}

To demonstrate the feasibility and usability of our AniVis approach, we build a web-based prototype system (Fig.~\ref{fig:poc}). The user interface of the prototyping system consists of a chart-loading panel with a simple chart attribute editor (Fig.~\ref{fig:poc}(b1)), a chart preview panel (Fig.~\ref{fig:poc}(b2)), an animation canvas (Fig.~\ref{fig:poc}(b3)), a transition configuration panel (Fig.~\ref{fig:poc}(b4)), and an export panel (Fig.~\ref{fig:poc}(b5)). 

As illustrated in Fig.~\ref{fig:poc}(a), the inputs of the system are two JSON files that contain essential information, such as an underlying data table and chart specifications, for the source and target charts, respectively. 
Although the current format of the JSON file is customized for the current system, all the necessary information for the input of the pipeline can be extracted and converted from the common visualization specifications (e.g., Vega-Lite specifications~\cite{satyanarayan2016vega}). More example input files are given in the supplementary material. 
In the given example, the source chart is the line chart with two series, whereas target chart is the column chart without the \textit{Brand} dimension. Once these two files are loaded, the static previews of the chart pair will be rendered with D3.js~\cite{bostock2011d3} and displayed in the top panel (Fig.~\ref{fig:poc}(b2)). The user can then specify the display options for each chart from the chart attributes panel, which controls the appearance of basic elements (i.e., axes and legend) of a chart (Fig.~\ref{fig:poc}(b1)). 

Then, the system analyzes the transition units between the two charts through the tree-based model construction for the data properties in the specifications and the transition identification modules for the tree and visual properties. With the identified transition units, the system orders them and synthesizes an animated transition between the two charts, which can be previewed on the animation canvas (Fig.~\ref{fig:poc}(b3)). The user can further configure the transition generation to refine the animation results in the configuration panel (Fig.~\ref{fig:poc}(b4)). 
Informed by the guidelines for developing animated data graphics authoring tools~\cite{thompson2020understanding}, the system provides the controls of staging strategy, duration, ease function, and animation effects. 
For instance, the system has the animation-based and fix-time modes for the user to control the staging and duration of the animated transition. For the motion of each transition unit, the control of the ease function is also provided to the user. The user can choose \textit{Slow in/Slow out} instead of the default \textit{Linear} setting to make the transitions more perceptually pleasing and effective than before~\cite{dragicevic2011temporal}. Moreover, different animation effects can be applied to convey individual same data change semantics as pluggable design components. For example, the removal of a series in the line chart can either fall out of the chart or disappear with a fade effect. Once the user is satisfied with the animated transition result, they can simply click a button (Fig.~\ref{fig:poc}(b5)) and export the animation as a video clip (MP4 or GIF file) for easy sharing (Fig.~\ref{fig:poc}(c)). 
To see how the system works in action, please refer to the supplementary video.

\section{Evaluation}
In this section, we evaluate AniVis in three forms: 1) examples to demonstrate the capability to satisfy different transition types in visualization; 2) a comparative study with the animated transitions generated by the latest automatic system and designers; and 3) expert feedback on its usability.

\subsection{Example Animated Transitions}
In this section, we demonstrate how the data-driven animated transitions generated by AniVis satisfy various types of transitions through a range of examples. 
We choose examples on the basis of the transition types in data visualization by Heer and Robertson.~\cite{heer2007animated}, namely, \textit{Filtering}, \textit{Ordering}, \textit{Timestep}, \textit{Substrate Transformation}, \textit{Visualization Change}, \textit{Data Schema Change}.
Our automatic approach aims to create animated transitions between two single-view statistical charts; therefore, the role of \textit{View Transformation} (movement of camera) is not covered. 
The collected cases show the utility of the defined data change units by handling the transitions for different roles of interactions in the visualization. 
More animated transition examples are available in the supplementary material. 

\begin{figure*}[t]
\centering
  \includegraphics[width=0.97\linewidth]{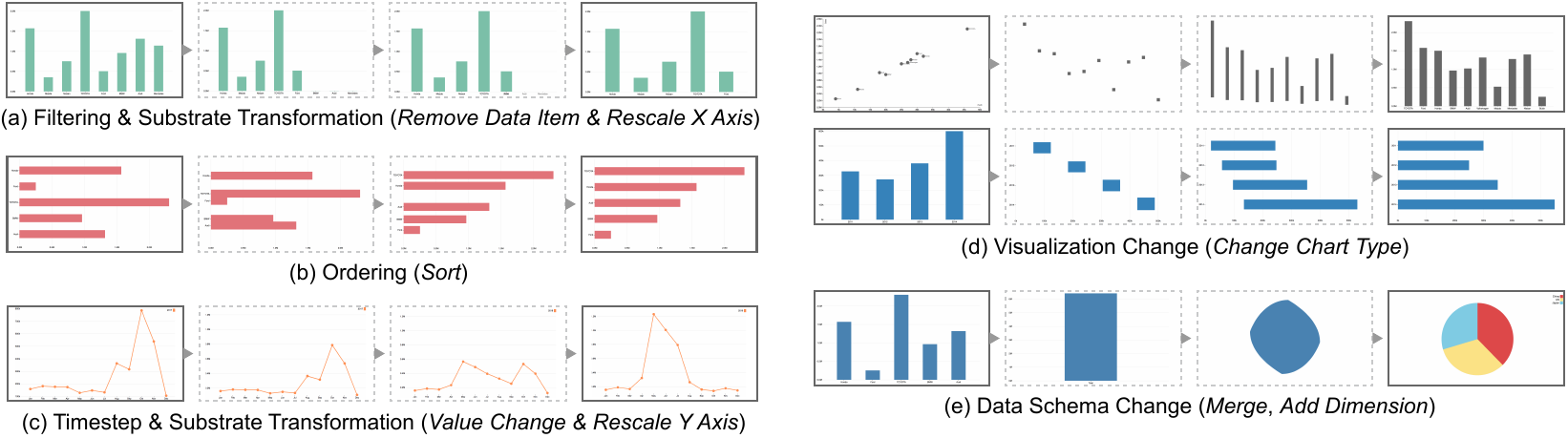}
  \caption{Examples of animated transition generated by AniVis: We see how animated transitions with different types of transition units can be generated successfully from source and target charts. Here, we hide chart titles for compact presentation. Additional examples can be found in the supplemental material.}
  ~\label{fig:gallery}
\end{figure*}

\textbf{Filtering.}
The set of the data items shown in the target chart can be filtered from those in the source chart. The animated transition of the filtering between the two charts can be achieved with the transitions for \textit{Remove/Add Data Item} unit in AniVis, which serve the purpose of conditional display. As shown in Fig.~\ref{fig:gallery}(a), the \textit{Remove Data Item} unit filters the brands in Germany in the bar chart. 

\textbf{Ordering.} 
The ordering transition that changes the spatial arrangement of representations is an essential operation to provide different perspectives on the dataset, which satisfies the user's intent to reconfigure for a different arrangement~\cite{yi2007toward}. Fig.~\ref{fig:gallery}(b) illustrates an animated transition between an original chart and an ordered one. AniVis automatically detects the data change unit of \textit{Sort} from the data subspaces of the two charts and identifies the corresponding data items between the charts to drive the new arrangement of the bars. Meanwhile, the names of the data items on the y-axis are moving at the same speed as their relative bars. 

\textbf{Timestep.} AniVis enables the generation of transitions for different timestamps. As shown in Fig.~\ref{fig:gallery}(c), when a temporal changes of data items is applied to the source chart, our approach can identify the data change unit of \textit{Value Change} and create animated transitions with the \textit{tweening} effect to reveal the changes of the data items. 

\textbf{Substrate Transformation.} Examples of substrate transformations include axis rescaling\cite{heer2007animated}. When data values change greatly or the number of data items change, AniVis inserts axis animations for rescaling in the final animated transitions. As shown in Fig.~\ref{fig:gallery}(a), the transition for the data-dependent unit of \textit{Rescale X Axis} is applied after the filtering transition. In Fig.~\ref{fig:gallery}(c), the animation for \textit{Rescale Y Axis} that adjust the scale of y-axis is inserted to fit the new range of data values. 

\textbf{Visualization Change.} 
AniVis also enables the change of the visual appearance, including the shape and color of the data elements, which support the user's intent to encode a different representation.  The upper part of Fig.~\ref{fig:gallery}(d) shows snapshots of the \textit{Chart Type Change} transition, where the scatter plot is transformed into a bar chart. The x-axis is first removed by using the \textit{fade out} effect, and the point marks are \textit{move} to the correct positions to align with the items in the bar chart. Then, the point marks gradually extend to line marks, and expand the width of the lines is expanded to turn the lines into bars. This visual change unit of \textit{Morphing} is designed to maintain valid data graphics and improve tracking~\cite{heer2007animated}. In the final stage, the axes are updated accordingly. 
Moreover, by extending the chart types and specific shape interpolation functions, AniVis can easily support the low-level shape morphing between the standard chart and the variant one (e.g., vertical bar chart to horizontal bar chart); see the lower half of Fig.~\ref{fig:gallery}(d). 

\textbf{Data Schema Change.} 
The data change units of \textit{Remove/Add Dimension} and \textit{Merge/Split Data Item} detected between the source and target charts are the novel ways to alter the level of abstraction of data representation. 
The example in Fig.~\ref{fig:gallery}(e) demonstrates how the animated transition bridges the different levels of details of individual data cases. The bars are first merged into one, which corresponds to the \textit{Merge Data Item} transition unit. Then, the shape of the merged data item is morphed from a rectangle to a circle, laying a foundation for the transformation to the pie chart in the target chart. Here, the circular mask represents the total values of the data items. Finally, the \textit{Add Dimension} transition unit fills the colors for the pie chart, thereby altering the representation from an overview down to details for a new category. With AniVis automatically identifying the data change units, users do not have to consider the intermediate steps or to design complicated animation effects to construct the transition (e.g., drill-down) that shows different levels of data abstraction. 

\subsection {Comparative Study}

To explore the advantages of our automatic approach, we performed an in-lab user study comparing our animated transitions with the results generated by the latest method~\cite{kim2020gemini} and designers. 
The objective was to determine the capability of our approach to assist the viewer in understanding the data changes between two statistical charts. 

\subsubsection{Study Preparation} 
To compare with the animated transitions generated by Gemini and designers, we took advantage of the stimuli in the replication study by Kim et al.~\cite{kim2020gemini}. 
To create the animated transitions for the stimuli with our method, we adapted the data source files provided on their website to our system and adjusted the style of the charts to match the stimuli. 
In addition, each leaf node in the current tree-based model assigns an aggregated value, limiting the support for data aggregation or disaggregation operation. We extended the tree model for the scatter plot and added an extra level to the leaves of the tree to represent the raw values. Then, the non-aggregated values can be supported by incorporating transition units of \textit{Aggregate Data Item} and \textit{Disaggregate Data Item}. However, the discrete tree model in the framework does not allow continuous and infinite dimensions. Thus, we omitted the case of filtering points with scatter plot and adopted the remaining cases that relate to our transition units: Expanding Lines (\textit{Add Data Item}), Sorting \& Updating Bars (\textit{Remove Measure}, \textit{Add Measure}, \textit{Sort}), and Aggregating (\textit{Aggregate Data Item}, \textit{Change Chart Type}), covering five of the seven transition types in the taxonomy by Heer et al.~\cite{heer2007animated}. 
Each case includes two animated transitions recommended by Gemini (i.e., single-stage design and best two-stage design) and two animated transitions that were manually authored by different designers. All stimuli used in the study can be found in the supplementary material accompanying this submission.

\subsubsection{Participants} 
We recruited 15 graduate students (6 females, 9 males) with a background in computer science and visualization (denoted as S1 to S15). They all have experience in conducting data analysis and creating visualizations by using tools such as Excel, D3, and Tableau. None of the participants has a visual impairment in viewing motion graphics. They confirmed that they have no prior knowledge of the dataset and stimuli used in our study. 

\subsubsection{Procedure}
The within-subjects study began with an introduction to the study goal and the study settings. 
In the following phase, we provided a chart pair to the participants and asked them to briefly describe the data changes between the two charts. 
Then, the participants were requested to perform a ranking task for the different animated transitions of the chart pair on the basis of how easy it is to identify and understand the underlying data changes. 
The participants were also required to replay all the animations and provide reasons for their rankings with at least three sentences from the perspectives of the ranking criteria, the reasons for the best and the worst, and consideration on design. 
Specifically, three chart pairs were used in this study, and each chart pair had five animated transitions generated by different methods for comparison. To ensure a fair comparison among the animated transitions, the orders of the animated transitions were randomized. 

\begin{figure}[tb]
\centering
  \includegraphics[width=0.97\columnwidth]{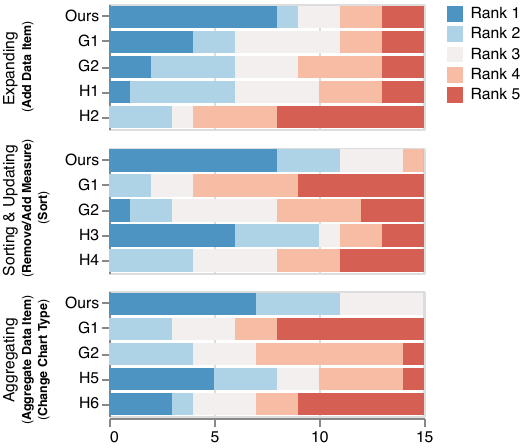}
  \caption{The participants' comparison rankings of animated transition generated by different methods (Ours: animations generated by AniVis; G1: animations best recommended by Gemini with a single stage; G2: animations best recommended by Gemini with two stages; H1, H2, H3, H4, H5, H6: The hand-crafted animations by different designers in the study by Kim et al.~\cite{kim2020gemini}). Note that Rank 1 is the best, whereas Rank 5 is the worst.}~\label{fig:comparison_ranking}
\end{figure}

\subsubsection{Results}
The participants took an average of 27.9 (SD = 10.8) minutes to finish all the tasks. 
Fig.~\ref{fig:comparison_ranking} shows the overall distributions of participants' rankings for different animated transition designs. We find that the designs generated by our approach obtain better results than the others (the highest-ranked designs on average for the three stimuli). We further compare the transitions by examining qualitative responses and use the Friedman test to determine whether a significant difference exists among the animated transition designs of each chart pair. 

For the stimulus of aggregation, we see a significantly stronger preference ($p < 0.01$) for our design over the single-stage design (G1) recommended by Gemini among the participants. 
Participants found our design helpful in depicting the data changes compared with the others. One participant (S8) commented that ``\textit{(Transitions generated by AniVis) clearly shows where the `average data' is derived from}.'' The results are credited with the separated data change units detected from the data transformation between the two charts with our approach. 
What's more, the ranking results of G1 and G2 indicate that the fading effect in the animated transitions offers little help in understanding the data change process. Another participant (S12) explained, ``\textit{Example E (by AniVis) shows that each value in the second chart is derived from the mean in a clear, visual way that removes ambiguity...Example C (by Gemini) is smooth with a nice fade. But it has the smallest amount of animations, which are very helpful to draw the attention of observer and show the details}.'' 

However, the results are not significant ($p > 0.05$) in other stimuli, indicating that the animated transitions generated by AniVis are not always significantly better than the others in facilitating the understanding of data changes between charts. For the stimulus of expanding lines with simple changes in data, animated transitions obtain a better ranking because of the fluency and simplicity. We speculate that difference among the transitions is small when the changes in data are uncomplicated. 
One participant (S3) expressed, ``\textit{Option B (by AniVis) has a nice idea, but it's more confusing than both A (by Designer 2) and C (by Gemini with single stage) since they more `compact'},'' suggesting a more advanced grouping strategy in our pipeline when multiple steps occurs in a relatively short time. The number of the top ranked for G1 also reflects the point of view. 
For the stimulus of sorting and updating bars, human design is sometimes better than ours from the point of view (S7) that ``\textit{Clearing the data chart and changing the y-axis information before shifting the data columns gives a clearer view of the change}.'' 
Although the staged units in our transitions prevent overlapping and help the viewer keep track of the bars, participants may have different preferences on the order of the transition units.

\subsection{Expert Feedback}
To evaluate the usability of AniVis and the quality of its generated transitions, we also collected feedback from four experts who have never used our system before the interview. 
Two of them (R1, R2) are animated visualization researchers who are familiar with animation design principles, and the remaining two (A1, A2) are digital media practitioners with more than three years of experience in dynamic multimedia content. 

The interview has two stages. In the first stage, we provided the experts with two source and target charts. They were asked to 1) describe their working processes and the detailed steps for generating a desired animated transition between these two charts and 2) assess how much effort is needed to accomplish the transition. 
In the second stage, we gave a 15-minute introduction of AniVis and demonstrated our prototype system. Then, we asked the experts to freely explore our system and generate animated transitions with their own configurations to further understand how AniVis is used in real-world scenarios. After the exploration, we asked the experts to comment on the usability of the tool and the quality of the results. Finally, we collected their suggestions in terms of their additional requirements. The whole interview took approximately 60 minutes. 

Overall, the experts valued the usefulness and efficiency of the system. The two designers were impressed by the results automatically generated. When we explained the basic pipeline of the system to the experts, they soon understood and agreed with the pipeline. Compared with their typical creation processes, using either programming languages or professional tools, the experts expressed that the system is convenient to generate animations with few clicks and would be helpful for data presentations. 

\paragraph{Feedback on the System.}
A1, who has more than six years of expertise in digital media, emphasized that ``\textit{Although the professional tools, such as C4D, are very powerful, I have to identify the difference and consider the mappings in animation. It will take me around one or two days to build one from scratch.}'' 
A2 and R1 also faced a similar labor-intensive situation as A1. They have more than three years of experience generating digital media with p5.js or D3.js, and they appreciated the overall design of the workflow because it greatly saves their time and effort managing the data transitions by their code. R1 said she must spend time figuring out the data changes before writing code for the animated transition.  ``\textit{The pipeline directly detects the data changes between two data visualizations, which also improve the efficiency of the whole creation process,}'' R1 added. 
Given the design background, A2 thought it can be used for the rapid drafting of sophisticated animation between statistical charts. The designer valued that ``\textit{It offers the opportunities for the re-creation of data-related animations.}'' Hence, he considered it as an easy-to-use sketching tool and made further suggestions on providing additional output options for future editing. 

\paragraph{Feedback on the Generated Results.} 
Regarding the quality of the generated animated transitions, the designers commented that the outputs are professional enough for most uses. A1 said, ``\textit{The final animation is basically the same as I planned.}'' 
``\textit{Similar to the auto-generated results, I will group the same type of transition units to save time and indicate the relevant relationships of the visual elements. I also tend to remove the undesired elements first, and then add the new ones later.}'', commented R2. 
For the extra configurations we have provided, such as duration, ease function, and the effect options for a certain transition unit, the experts felt that they are flexible and sufficient for typical cases, saying ``\textit{The tool is obviously not for producing elaborate animations with complex effects; thus, keeping it clear and efficient is important.}''

\paragraph{Suggestions. } 
Apart from all the positive feedback of the resulting transitions, the experts and designers also made a few suggestions. For example, A1 said, ``\textit{Movements in real life do not start or end in a perfectly synchronized manner. I think allowing actions to overlap can make the movements more realistic.}'' R1 and A2 also thought that displaying a timeline and allowing customized timing control for specific items can improve attraction and engagement. For instance, A2 expressed that emphasizing on the elements to be moved and applying some artistic acceleration effects on the movement would be interesting. 

\section{Discussion}
\textbf{Automatic Generation of Animated Transitions.} 
AniVis makes it possible for non-designers to create animated transitions for data changes between charts with the automatic composition of animations. The experts in our user study said the system is easy to learn and enables the rapid generation of various data animations between different data views. 
Comments from the user study suggest that the animated transitions generated by our framework can help users identify the critical changes in data transformations. Thus, the generated animations might be integrated into the process of data analysis, enabling users to track changes and receive feedback about their analysis. 
Moreover, users can start with a good draft with the generated animations instead of starting from scratch. One designer from the expert interview stated that animations could be produced with little user intervention, and considered the system as a prototyping tool for his commercial data video creation. In addition to the animated transitions created with default configurations, the system provides necessary controls to help users refine the visual effect of the generated results. 

\textbf{Trade-off Between Automation and Control.} 
AniVis demonstrates the feasibility and capacity of modeling an animated transition by breaking down the process into operational units. However, a trade-off always exists between the automatic pipeline and customized control. In our current proof-of-concept implementation, users have the flexibility to select the staging strategy and several basic dynamic properties to control the motions of transition units. However, the order of the transition units is automatically computed based on the tree models of the charts. During the open-ended discussion, one expert raised a question about whether the sorting function can be moved ahead before the chart type change process. However, the current automatic pipeline may not guarantee optimized results if we allow users to randomly reorder the animation sequence. An alternative way to alleviate the issue is to provide multiple feasible animation sequences for the users to meet different requirements. The trade-off also indicates research opportunities for animation sequences. For example, how do different sequence choices affect subsequent decision-making or information recall? In addition, the ``correct'' sequence for a given situation may vary with contexts, such as the dataset, the intended audience, and the presenter's goals. 

\textbf{Animation Authoring Tools.} 
In this paper, we demonstrate AniVis as a standalone system that solely targets animated transitions. It is possible to convert the charts from other charting systems (e.g., Vega-Lite~\cite{satyanarayan2016vega} and Charticulator~\cite{ren2018charticulator}) into ours to improve the rendering of the charts in the final animated transitions. Meanwhile, additional types of visual properties, such as style and annotation, need to be considered for extending the grouping and ordering modules in our pipeline. 
Although the current system is developed for the basic chart types, it is extensible for more complex cases. For example, bubble charts and radial stacked bar charts are essentially the visual representation variants of the standard stacked bar chart from the tree-based data model perspective. Thus, the new chart types can be feasibly introduced by implementing specific animation effects without changing the underlying transition generation workflow. We would like to extend various chart types as well as their animation effects in future work.
The interview session indicates that the designers hope that more different animation effects could be applied as pluggable components under the current framework. Hence, we can enhance the expressiveness of the resulting transitions by incorporating the templates for the staging and visual appearance, which provide rich styles (e.g., exaggeration) for motion effects. 

Our auto-generation method also suggests new opportunities for data video authoring. A data video often contains a series of semantically related data views. Although users can generate a list of animated data transitions independently by using AniVis with little user intervention, directly combining them into a data video may not achieve desired results. Additional considerations, such as the consistency between each data view and the emphasis of the insight, are required by the data video authoring system.

\textbf{Limitations.}
The approach we have developed represents an attempt to automatically generate animated transition driven by the underlying data relations between statistical charts. Thus, the approach has limitations arising from our assumptions that helped simplify the creation process. 

In this paper, we focus on data views with only a single chart. The current framework cannot support data views with multiple charts, which is commonly seen in a data story or a data report in practice. One potential workaround is to achieve one-to-one chart correspondence between scenes. Then, the transition between the two data views can be treated as a set of single chart transitions. However, a more in-depth investigation of the relationships among the charts represented by the tree-based model is needed to fully support more complicated scenarios. It would be an interesting direction for future work. Moreover, we currently assume that the transition of each visual element is independent from one another, which may lead to unsynchronized results. For instance, when animating a visual component with a text annotation, the movement of the text and the visualization part cannot be arranged cohesively. Therefore, a more refined time control is desired to improve the output of the system. 
As mentioned in the previous section, the tree-based data model in our pipeline can only handle discrete values, which means it has a limitation for continuous and infinite dimensions. It is interesting to support this kind of data field, and further exploration is needed to extend the current framework. We encourage follow-up research efforts that investigate the effective animated transitions for smooth data fields.

In addition, the transition units generated by our system are generally staged and separated on the basis of the \textit{Staging} principle proposed by Fisher~\cite{fisher2010animation}. According to the comments from the open-ended discussion, complex transitions may generate a long animation sequence, which is boring if the duration is too long or difficult to track if the time length is limited for viewers. Although we have mechanisms to prevent complex transitions (e.g., use simple animation effects, group similar transition units, limit the number of data transformations, etc.), the current results can be further optimized. For future work, we intend to develop a more intelligent simplification method that minimally disturbs the perception of changes (e.g., detect the latent changes in data and reduce redundant animation steps) and conduct a deeper investigation on maintaining the understandability of the animated transitions without increasing the complexity and consuming too much time. 

\section{Conclusion and Future Work}
We present AniVis, a systematic approach that leverages the tree-based model for data to automatically generating animated transitions from static charts. AniVis first models the data and visual properties of data charts and formalizes the differences between the two charts with a set of data and visual change transition units. Through a composition process, AniVis assembles the transition units into a sequence. Subsequently, AniVis corresponds the transition units to a series of default animation effect designs from the animation effect library, and the animated transitions are produced. 
We also develop a prototype system and demonstrate the usage of the generated transitions with a series of examples. Through a user study and expert interviews, we show that our approach can create intuitive and usable transitions with ordered transition units, which give a clear view of data changes and facilitate the creation process of user-desired animations. 
In the future, we will support other chart types (e.g., bubble, area chart, and donut charts) by considering the variations of the current standard ones and will extend our animation transition effect library with other styles. Furthermore, investigating the precise priorities and timing of different transition units, which allow us to model the sequencing of multiple animations, is also an interesting direction. 

\begin{acks}
The authors thank the anonymous reviewers for their valuable comments. 
The authors also thank all the participants for their participation in the user studies and interviews.
\end{acks}

\balance
\bibliographystyle{ACM-Reference-Format}
\bibliography{reference.bib}

\end{document}